\documentclass{ifacconf}

\makeatletter
\let\old@ssect\@ssect %
\makeatother

\usepackage{graphicx}      %
\usepackage{svg}
\usepackage{natbib}        %
\usepackage{amsmath}
\usepackage{amssymb}
\usepackage{siunitx}
\usepackage{hyperref}
\usepackage[utf8]{inputenc}
\usepackage{multirow}
\usepackage{textcomp}

\usepackage{fancyhdr}
\fancypagestyle{firstpage}{%
  \fancyhead{}
  
  \fancyfoot[C]{\small{\\\textsuperscript{\tiny{\textcopyright}} 2025 Cedric Bös, Alessandro Bortotto, Prof. Mohamed Khalil Ben-Larbi.\\
  This work has been accepted to IFAC for publication under a Creative Commons Licence CC-BY-NC-ND.}}
}

\makeatletter
\def\@ssect#1#2#3#4#5#6{%
  \NR@gettitle{#6}%
  \old@ssect{#1}{#2}{#3}{#4}{#5}{#6}%
}
\makeatother

\newcommand{\angstrom}{\mbox{\normalfont\AA}}   %
\begin{document}
\begin{frontmatter}

\title{Forecasting Thermospheric Density with Transformers for Multi-Satellite Orbit Management}

\author[Affiliation1]{Cedric Bös} 
\author[Affiliation1]{Alessandro Bortotto} 
\author[Affiliation1]{Mohamed Khalil Ben-Larbi}

\address[Affiliation1]{Julius-Maximilians-Universität Würzburg, Chair of Space Computer Science and Satellite Systems, Am Hubland, Würzburg, 97074, Germany (e-mail: \href{mailto:cedric.boes@stud-mail.uni-wuerzburg.de}{cedric.boes@stud-mail.uni-wuerzburg.de})}

\begin{abstract}                
Accurate thermospheric density prediction is crucial for reliable satellite operations in Low Earth Orbits, especially at high solar and geomagnetic activity.
Physics-based models such as TIE-GCM offer high fidelity but are computationally expensive, while empirical models like NRLMSIS are efficient yet lack predictive power.
This work presents a transformer-based model that forecasts densities up to three days ahead and is intended as a drop-in replacement for an empirical baseline.  
Unlike recent approaches, it avoids spatial reduction and complex input pipelines, operating directly on a compact input set. 
Validated on real-world data, the model improves key prediction metrics and shows potential to support mission planning.
\end{abstract}

\begin{keyword}
time series, thermospheric density, satellite data, artificial intelligence, transformer
\end{keyword}

\end{frontmatter}

\thispagestyle{firstpage}

\section{Introduction}
\vspace{-0.3cm}
The unprecedented growth in satellite deployments, particularly by large constellations in Low Earth Orbit (LEO), has intensified concerns over orbital crowding, increasing the frequency of close approaches and potential collisions~\citep{BenLarbi.2022}. 
These developments have elevated conjunction risk assessment and collision avoidance to operational necessities for spacecraft (s/c) operators. 
However, the effectiveness of these safety-critical measures depends fundamentally on the ability to accurately model and predict the spatio-temporal evolution of the resident space object (RSO) population. 
This capability relies on several key factors, including precise object tracking, adequate sensor coverage, reliable uncertainty propagation, and accurate modeling of environmental perturbations such as atmospheric drag,  which is primarily governed by thermospheric density. 
Even more critical, atmospheric drag is not only a perturbation to be modeled but also a control input actively exploited in modern mission concepts for propellantless maneuvers, including formation flight~\citep{BENLARBI20213444}, and collision avoidance~\citep{TRAUB2025296} through differential drag modulation.
While progress in tracking, coverage, and uncertainty modeling has improved short-term conjunction prediction, forecasting thermospheric density remains a critical limitation. 
Uncertainties in density modeling can lead to orbit prediction errors of tens of kilometers within days, yet robust tools for estimating future orbit density under realistic assumptions about object propagation, atmospheric drag, solar activity, and satellite operations remain lacking. 
Thermospheric density is primarily influenced by temperature variations caused by solar activity~\citep{knipp2004direct}, particularly due to the absorption of extreme ultraviolet (EUV, $30-120 \si{\nano\metre}$) radiation, and the complex interaction between solar wind particles and Earth's geomagnetic field leading to Joule heating~\citep{qian2012thermospheric}.
These effects are especially pronounced during periods of high solar activity, a phase that can currently be observed~\citep{petrovay2020solar, sawadogo2024solar}. 
Recent space weather events, such as the May 2024 Gannon geomagnetic storm, have underscored the vulnerability of LEO operations to density forecasting errors~\citep{Parker.2024}.  
During this event, drag-induced orbit decay increased by a factor of four for many satellites, triggering thousands of unplanned maneuvers. 
Automated station-keeping systems, particularly within large constellations such as Starlink, responded en masse, leading to inter-satellite phasing disruptions, overwhelmed conjunction screening pipelines, and invalidated predicted ephemerides~\citep{Parker.2024}. 
These operational disruptions reveal the urgent need for robust predictive tools that go beyond static models or naïve persistence-based forecasting. 
Traditionally, thermospheric density is modeled using either physics-based approaches such as TIE-GCM~\citep{qian2014ncar}, or empirical models like NRLMSIS-2.1~\citep{emmert2022nrlmsis} and JB2008~\citep{bowman2008new}.
Physics-based models offer high physical fidelity by solving the coupled continuity, momentum, and energy equations for the neutral and ionized upper atmosphere. 
However, they are computationally expensive, sensitive to initial and boundary conditions, and difficult to run in real-time or onboard settings.
As a result, empirical models remain the operational default due to their speed and simplicity. 
Yet, they rely entirely on historical data and often fail to reproduce the chaotic variability of space weather processes.
This limitation becomes especially problematic during solar maximum, when EUV irradiance and geomagnetic storms drive abrupt non-linear changes in density~\citep{knipp2004direct, qian2012thermospheric}.
To address these challenges, recent work has explored Deep Learning (DL) methods. 
These models aim to combine the computational speed of empirical models with learned representations of spatiotemporal dynamics, offering promising generalization capabilities even under unmodeled solar conditions.
Earlier studies have already demonstrated the great potential of simple neural networks and long short-term memory (LSTM) networks for space-weather time series forecasting~\citep{camporeale2019challenge}.
Building on this success, \citet{turner2020machine} use autoencoders to better encode highly nonlinear thermospheric density in reduced-order models, and deep neural networks to improve their predictions.
A transformer model developed by \citet{briden2023transformer} to propagate thermospheric density showed great performance compared to linear models such as DMDc.
\citet{pelkum2024forecasting} also use a transformer architecture to successfully predict the release of solar flares in coming days.
Recently, \citet{li2025prior} proposed using a hybrid Resnet architecture with prior knowledge of NRLMSIS-2.1 to predict thermospheric density under quiet geomagnetic conditions, which converged quicker and exhibited better generalization performance than a default Resnet.

Motivated by this potential and in response to the \textit{2025 MIT ARCLab Prize for AI Innovation in Space}, this work presents a fast and effective transformer-based surrogate model to forecast thermospheric density\footnote{https://github.com/cb0s/ifac-camsat-2025-timeseries-transformer}.
Inspired by the success of \citet{li2025prior}, the proposed model takes an empirical baseline, i.e. NRLMSIS-2.1, as an additional input, leading to faster and more reliable convergence.
We also test this model in a residual setting, where instead of predicting absolute density values, the model forecasts the residual between the observed density and that predicted by an empirical baseline model, effectively learning to correct systematic errors over time.
We speculate that the residual learning approach simplifies the learning task and improves the robustness to baseline model drift.
The remainder of this paper is structured as follows:
\autoref{sec:methodology} describes the model architecture, training strategy, and input feature design. 
\autoref{sec:results} presents experimental results and performance benchmarks. \autoref{sec:discussion} discusses operational implications and limitations, and \autoref{sec:conclusion} concludes with final remarks and future outlook.

\section{Methodology}
\label{sec:methodology}
\vspace{-0.2cm}
The objective is to develop a model $\theta(x)$ that, given a compact set of inputs $x$, predicts the thermospheric density $y_t$ over a three-day forecast horizon at 10-minute intervals ($t \in [1, 432]$) for a specific RSO with known initial orbital elements. 
In this work, we evaluate two training strategies based on the same model architecture: one predicting residuals with respect to the empirical baseline, and one using an end-to-end formulation to directly forecast density values.
\autoref{tab:compared-models} summarizes the key differences.\vspace{-0.2cm}

\begin{table*}[!ht]
\centering
\caption{Overview of the compared models with their respective input-output configurations.
We compare the persistence baseline model against our transformer-based model trained using two strategies: one predicting residuals with respect to the baseline, and one trained directly on measured density values.}
\label{tab:compared-models}
\resizebox{0.85\textwidth}{!}{
\begin{tabular}{|l|c|cc|}
\hline
\textit{Model} & \textbf{Persistent NRLMSIS-2.1} & \multicolumn{2}{c|}{\textbf{Timeseries Transformer}} \\ \hline
\textit{\begin{tabular}[c]{@{}l@{}}Training\\ Method\end{tabular}} & n./a. & \multicolumn{1}{c|}{residual approach} & end-to-end approach \\ \hline
\textit{Inputs} &
\begin{tabular}[c]{@{}c@{}}time and location information,\\ proxy indices for approximating\\ sun-induced flux and magnetic field\end{tabular} &
\multicolumn{2}{c|}{\begin{tabular}[c]{@{}c@{}}99 hand-selected inputs that are mostly shared with\\ either the NRLMSIS-2.1 or TIE-GCM and its dependencies\\ (more details can be found in \autoref{tab:dataset-info})\end{tabular}} \\ \hline
\textit{Outputs} &
\begin{tabular}[c]{@{}c@{}}simulated\\ value at t=0\end{tabular} &
\multicolumn{1}{c|}{\begin{tabular}[c]{@{}c@{}}residuals to\\ persistence baseline\end{tabular}} &
\begin{tabular}[c]{@{}c@{}}measured ground\\ truths of satellites\end{tabular} \\ \hline
\end{tabular}
}
\end{table*}

\subsection{Data}
\vspace{-0.2cm}
The challenge provided two key input data sources: solar activity data from the GOES-EAST satellites and the OMNI2 collection maintained by NOAA~\citep{Papitashvili2020}.
GOES data consisted of minute-resolution X-Ray measurements from two spectral channels ($0.5-4\angstrom$ and $1-8 \angstrom$).
OMNI2 is a composite data set that aggregates solar wind and magnetic field parameters, plasma data, and geomagnetic indices from multiple spacecraft, available at an hourly cadence. 
Ground truth density values were derived from in-situ accelerometer data from the SWARM-A, CHAMP, and GRACE-2 missions.
Additionally, own orbit simulations based on initial conditions from those satellites (six orbital elements and time) were carried out, yielding entry and exit angles for the umbra and penumbra for each object.
To encode time-dependent features, we first scale the values to be $2\pi$-cyclic followed by sinusoidal encoding, which effectively captures cyclic patterns and is easier for Neural Networks to learn.
The data is structured into overlapping three-day windows, each offset ranging from hours to more than one day, resulting in $8119$ raw samples.

Missing values were handled differently depending on the type of data. 
Ground truth gaps were treated conservatively, as a consistent target signal is essential for supervised training.
Interpolation was applied only if the statistical moments (mean, standard deviations) remained stable and extrema values unchanged; otherwise, the sample was discarded. 
This filtering step reduced the dataset to approximately $7700$ samples. 
In contrast, missing values in the input features were addressed through a multi-resolution downsampling strategy.
Each feature was aggregated using the maximum, mean, and standard deviation over a subset of time windows of 3 hours, 24 hours, 3 days, 14 days, and 60 days. 
These intervals correspond to meaningful physical time scales: twice the orbital period of LEO satellites, Earth's daily cycle, the prediction horizon, half the solar rotation period~\citep{sonett1991sun}, and the smoothing window used as a proxy for the F10.7a index, respectively.
Minimum values were excluded from aggregation, as density fluctuations are typically dominated by extreme solar events, which are better captured through maxima.
Any remaining missing inputs were treated as ``natural dropout'', analogous to dropout regularization in deep learning.
In essence, this means that these values were zeroed and the model is trained to not solely rely on one input alone, and thus naturally learns to deal with missing values.
This technique proved to be effective, enabling training even on sparse datasets, a common problem in space weather applications.
For measures that were nearly constant, i.e., whose minimum and maximum were almost identical, they were dropped because we believe that they can be learned by the model directly.\\
With the curated data, we used NRLMSIS-2.1 to simulate the thermospheric density at the time of generation $t_0$ and used this as a perstistence baseline where $y_{t} = y_{t_0}$.
This value is encoded using $\log(\cdot)$ and is used as an additional input for the model.

The input fields were chosen to be similar to those of the NRLMSIS-2.1 baseline model and the TIE-GCM and its respective dependencies.
These proxy values in our opinion describe the situation to be simulated well, hence a good model should be able to reflect this dependency.
The additional eclipse parameters from the orbit simulations are supposed to simplify the assessment of the influence of solar wind onto the experienced atmosphere by the s/c.
Recent work also found that adding a baseline value to start from can improve the performance and increase the speed of convergence of the subsequent model~\citep{li2025prior}.
\autoref{tab:dataset-info} provides a complete list of aggregated features with descriptions.

\begin{table*}[!ht]
    \centering
    \caption{Description of the input fields used for the proposed transformer model. This model uses $99$ input fields in total. Each of the time aggregations, which are indicated by brackets, are performed using $\text{max}$, $\text{mean}$, and standard deviation ($\text{std}$), except if there is only a single value.}
    \resizebox{0.88\linewidth}{!}{
    \begin{tabular}{|l|c|l|c|}
    \hline
    \textbf{Field name(s)}                                                                                                  & \textbf{Original Dataset} & \textbf{Description}                                                                                                                                                             & \textbf{Aggregation} \\ \hline
    \begin{tabular}[c]{@{}l@{}}Altitude, Inclination\\ RAAN\end{tabular}                                                    & Initial States            & \begin{tabular}[c]{@{}l@{}}Part of initial orbital elements\\ determines the position in orbit\end{tabular} & [single value]          \\ \hline
    \begin{tabular}[c]{@{}l@{}}Orbit Angle\\ (True Anomaly +\\ Argument of Perigee)\end{tabular}                            & Initial States            & \begin{tabular}[c]{@{}l@{}}Part of initial orbital elements\\ determines the position in orbit\end{tabular} & [single value]          \\ \hline
    F10.7, Kp, Dst Indices                                                                                                  & OMNI2                     & \begin{tabular}[c]{@{}l@{}}Proxies for capturing thermospheric heating\\ and circulation, and geomagnetic disturbances\end{tabular}                                              & [14d, 3d, 3h]        \\ \hline
    60-day average F10.7                                                                                                    & OMNI2                     & \begin{tabular}[c]{@{}l@{}}Proxy for F10.7a for capturing long term trends in\\ thermospheric heating\end{tabular}                                                               & max, std, mean       \\ \hline
    Lyman-$\alpha$                                                                                                          & OMNI2                     & \begin{tabular}[c]{@{}l@{}}Proxy for capturing upper thermospheric composition\\ and heating\end{tabular}                                                                        & [14d, 3d, 3h]        \\ \hline
    \begin{tabular}[c]{@{}l@{}}Proton density,\\ Plasma speed\end{tabular}                                                  & OMNI2                     & \begin{tabular}[c]{@{}l@{}}Proxies for capturing the effect of sun wind onto the\\ magnetic field\end{tabular}                                                                   & [14d, 3d, 3h]        \\ \hline
    By, Bz                                                                                                                  & OMNI2                     & Capturing effects of the geomagnetic field                                                                                                                                       & [24h, 3h]            \\ \hline
    \begin{tabular}[c]{@{}l@{}}Solar cycle (Year / 11),\\ Day-of-year, Hour-of-day,\\ Bartel's Rotation number\end{tabular} & OMNI2                     & \begin{tabular}[c]{@{}l@{}}Capturing seasonal effects of the output,\\ all values are used with appropriately scaled Sinusoidal encoding\\ where  cycles are mapped to a $2\pi$ interval\end{tabular}                                                         & [single value]          \\ \hline
    \begin{tabular}[c]{@{}l@{}}Penumbra's and Umbra's\\ entry and exit orbit angle\end{tabular}                             & [Preprocessing]           & \begin{tabular}[c]{@{}l@{}}Helps to assess the direct influence of the solar wind\\ onto the s/c obtained with an orbit simulation with orekit\end{tabular}                      & [single value]          \\ \hline
    XRS-A, XRS-B                                                                                                            & GOES                      & \begin{tabular}[c]{@{}l@{}}X-Ray measurements of the GOES satellites from sensor\\ system A and B, help to quantify the influence of the high\\ energetic radiation\end{tabular} & [14d, 3d, 3h]        \\ \hline
    Persistence Baseline    & [Preprocessing] & Simulated values using NRLMSIS-2.1 & [single value] \\ \hline
    \end{tabular}
    }
    \label{tab:dataset-info}
\end{table*}

Due to small datasets and the high variability of thermospheric density, directly learning the absolute density values $y_t$ is challenging. 
To reduce the complexity of the learning task, in addition to end-to-end training, we also try to adopt a residual learning approach. 
Specifically, we used the value $y_b := y_{t_0}$ from the empirical persistence baseline model and trained our transformer to approximate the residual $\hat{y}_r = \hat{y}_t - y_b$, where $\hat{y}_t$ is the interpolated ground truth value. 
We believe that this formulation reduces the function space the model must optimize over, allowing it to focus on correcting systematic errors in the empirical baseline rather than modeling the full physical dynamics from scratch.

Finally, input features $\chi$ and the different targets $\hat{y}_t$ and $\hat{y}_r$ were standardized using z-score normalization to improve numerical stability and to support augmentation with Gaussian noise:
\begin{align}
x &= \zeta_x\left(\chi\right) \\
\bar{y} &= \zeta_y\left(\hat{y} \right) \\
\bar{y}_{r} &= \zeta_{y_r}\left(\hat{y}_r\right)
\end{align}
The dataset was divided into $80\%$ training and $20\%$ validation splits.

\subsection{Model}
\vspace{-0.2cm}
We propose a model based on the original Transformer architecture \citep{vaswani2017attention}, chosen for its ability to model long-range dependencies and scale effectively with increasing data volume and model complexity.
Unlike typical time-series transformers that leverage both endogenous and exogenous inputs~\citep{wu2021autoformer, wang2024timexer}, our setting is restricted to multivariate exogenous inputs only, as no measured atmospheric density observations are available during inference.

Despite the absence of autoregressive input, we maintain an encoder-decoder architecture.
The encoder ingests the full set of normalized inputs.
No patching or tokenization is applied, as we already work with a limited set of highly curated variables. 
Following the reasons outlined by~\citet{wang2024timexer}, we include cross-attention to efficiently align encoded input feature representations with the growing output sequence.
The decoder is initialized with a static start value of $0$, which serves as a pseudo-input to start the autoregressive generation.
The original sinusoidal encoding proposed by~\citet{vaswani2017attention} is used for the positional encoder.
To improve training stability, we apply pre-layer normalization (Pre-LN) before all attention and feed-forward blocks~\citep{xiong2020layer}.

The best performing model uses one encoder layer and one decoder layer, respectively, with an embedding dimension of $112$, a feed-forward size of $4 \cdot 112 = 448$, $4$ attention heads in each attention block, and $\text{Gelu}(\cdot)$ \citep{hendrycks2016gaussian} activation functions.
A small dropout of $0.1$ is added to reduce the chance of overfitting. \\
We compare this time series transformer to the aforementioned persistence baseline and train our model with both a residual target towards the persistence baseline and in an end-to-end approach simulating the ground truth directly.
A summarizing breakdown can be found in \autoref{tab:compared-models}.\vspace{-0.2cm}

\subsection{Training}
\vspace{-0.2cm}
The best performing Transformer was trained for $5000$ epochs (the best epochs were $1400$ and $4940$ for the end-to-end and residual training, respectively) using the AdamW Optimizer \citep{loshchilov2017decoupled} with a learning rate of $\alpha=\num{5e-5}$ and a weight decay of $0.01$, no early stopping, and a cosine learning rate scheduler with linear warm-up for $1\%$ of the epochs.
The learning rate scheduler was set to complete $35\%$ of one complete cycle, which effectively stops the training as soon as the learning rate hits $\frac{\cos\left(0.35 \cdot 2\pi\right)}{2} \alpha \approx \num{1e-5}$.
To foster better generalization of the final model, we used a low batch size of $32$~\citep{keskar2016large}, Gaussian-distributed input noise with $\text{std}=0.01$, and a very low output noise of $\text{std}=0.0005$, as this directly perturbs the training target.

For grading the models in the original MIT challenge, a custom metric called $\text{OD-RMSE}(y, y_b, \bar{y})$ was used that focuses more on the early predictions rather than on the complete forecasted sequence.
This metric is defined as:

\begin{align}
    \text{OD-RMSE}(y, y_b, \bar{y}) &= \frac{\sum_t w(t) \cdot \left(1 - \frac{\text{RMSE}(\bar{y}_t, y_t)}{\text{RMSE}(\bar{y}_t, y_b)}\right)}{\sum_t w(t)}
\end{align} 

where $w(t) = e^{-\gamma t}$ and $\gamma = -\frac{\ln {\epsilon}}{T}$ with $\epsilon = \num{e-5}$ as a numerical constant and $T$ being the total time of the predicted series, i.e., $T=432$.
However, it is defined on $\text{OD-RMSE} \in (-\infty, 1]$ and is supposed to be maximized.
To use it in training, it needs to be reformulated into a loss $\text{OD-RMSE Loss}(y, y_b, \bar{y}) := \Omega(y, y_b, \bar{y}) \in [0, \infty)$, which in the case of the residual predictor, can be simplified to $\Omega(y, \bar{y}_r)$:

\begin{align}
    \Omega(y, y_b, \bar{y}) &= \frac{\sum_t w(t) \cdot \left(\frac{\text{RMSE}(\bar{y}_t, y_t)}{\text{RMSE}(\bar{y}_t, y_b)}\right)}{\sum_t w(t)} \\
    \Omega(y, \bar{y}_r) &= \frac{\sum_t w(t) \cdot \left(\frac{\text{RMSE}(\bar{y}_{r_t}, y_t)}{\sqrt{\frac{1}{n}\sum \bar{y}_{r_t}^2}}\right)}{\sum_t w(t)}
\end{align}

Although early predictions are arguably more important than later ones, we still want a predictor that performs reasonably well on the whole sequence.
We thus combine $\Omega$ with a classic mean squared error ($\text{MSE}$) loss to get:\\
\begin{align}
    \text{Loss}(y, y_b, \bar{y}) = \Omega(y, y_b, \bar{y}) + \text{MSE}(y, \bar{y})
\end{align}

This combined loss function showed an improvement of about $10\%$ compared to a simple $\text{MSE}$ loss in terms of root mean squared error ($\text{RMSE}$) and $\Omega$ in our preliminary tests and is therefore used for all conducted experiments.

\section{Results}
\label{sec:results}
\vspace{-0.2cm}
We evaluate the trained models using both the end-to-end and residual learning strategies on the validation set.
In some instances, the unscaled, post-processed predictions produced physically implausible values. These included negative densities or extreme overshoots during periods of elevated solar activity.
To mitigate this issue, we applied clipping to constrain the outputs roughly within the physical bounds observed in the training data: $\text{min} = \num{e-14}$ and $\text{max} = 2\cdot \num{e-11}$.
This improved the final $\text{OD-RMSE}$ by around $4\%$ and $<1\%$ for the residual and for the end-to-end training, respectively, and yielded more physically consistent density predictions.

\begin{figure}[!ht]
    \centering
    \includeinkscape[width=0.9\linewidth]{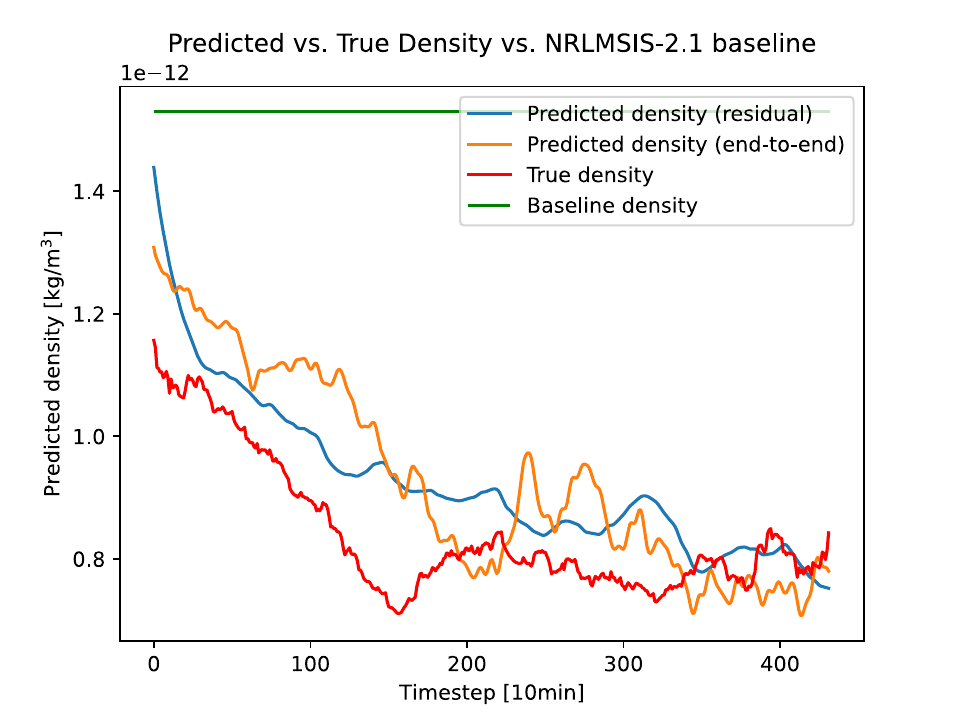}
    \vspace{-0.3cm} %
    \caption{
    Prediction of the thermospheric density using our proposed transformer with end-to-end and residual learning approaches compared to the persistence baseline and the measured density by GRACE-2 (July, 2013).\vspace{-0.15cm}
    }
    \label{fig:captured-well}
\end{figure}

As shown in \autoref{fig:captured-well}, the proposed transformer model deterministically outperforms the baseline in both training schemes.
In this instance, a moderate solar event occurred approximately two days prior to the prediction horizon, misleading the baseline forecast.
In contrast, our model with both training methods is capable of anticipating and correcting for this deviation.
Although the end-to-end strategy achieves lower absolute errors, the residual approach demonstrates smoother predictions over time.\vspace{-0.3cm}

\begin{figure}[!ht]
    \centering
    \includeinkscape[width=0.9\linewidth]{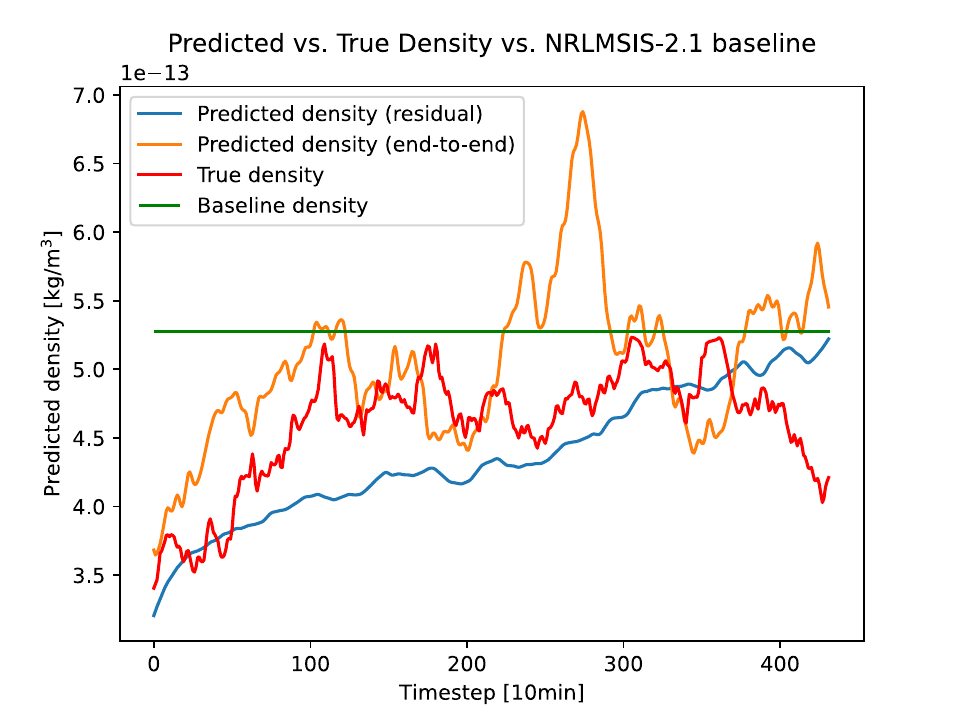}
    \vspace{-0.3cm} %
    \caption{
    Prediction of the thermospheric density using our proposed transformer with end-to-end and residual learning approaches compared to the persistence baseline and the measured density by SWARM-A (March 2016).
    Our transformer trained with both flavors is able to capture early events in rising densities.
    }
    \label{fig:predict-increase}
\end{figure}

In the instance shown in \autoref{fig:predict-increase}, the true thermospheric density increases as a result of a moderate solar event that occurred during the forecast horizon.
Although the baseline also captures a rise in density, due to its persistent nature, details about the lower initial conditions are lost.
In contrast, our transformer using both training schemes is able to model this increase.
The residual approach, by construction, acts as a bias-corrective model: instead of learning the full signal, it adjusts the empirical baseline.
This setup may promote smoother and more stable predictions.
The end-to-end approach, by comparison, is more reactive and captures finer fluctuations, which can occasionally lead to sudden and short hallucinated overshoots.\vspace{-0.15cm}

The model was evaluated against the persistence baseline using the mean absolute error ($\text{MAE}$), $\text{RMSE}$, mean absolute percentage error ($\text{MAPE}$), symmetric mean absolute percentage error ($\text{S-MAPE}$), and $\text{OD-RMSE}$.
To provide further insight into residual errors $\xi$, we also report the mean $\bar{\xi}$, standard deviation $\sigma_\xi$, and maximum absolute error $\max_\xi$.
As shown in \autoref{tab:results}, the proposed Transformer model significantly outperforms the baseline on all metrics.
In addition, residual statistics indicate improved consistency in prediction accuracy. 
Interestingly, the end-to-end training strategy generally achieves better overall performance. 
However, the residual approach yields lower errors in terms of relative metrics such as (S-)MAPE, likely due to its smoother output behavior. 
This suggests that the residual model is better suited for periods of low solar activity with smaller variations in density, whereas the end-to-end approach handles larger dynamic ranges more effectively.

\begin{table}[!ht]
    \centering
    \caption{
    Comparison of evaluation metrics and residual error statistics for the persistence baseline (NRLMSIS-2.1) and our proposed Transformer model. 
    Arrows indicate whether higher ($\uparrow$) or lower ($\downarrow$) values are better.
    }
    \begin{tabular}{|c|c|cc|}
    \hline
    \multirow{2}{*}{\textit{Metric}} & Persistent & \multicolumn{2}{c|}{Our Model}                                     \\ \cline{3-4} 
                                     &  NRLMSIS-2.1                            & \multicolumn{1}{c|}{residual}          & end-to-end                \\ \hline
    \textit{OD-RMSE} $\uparrow$      & $0.0$                        & \multicolumn{1}{c|}{$0.802$}           & $\mathbf{0.826}$          \\ \hline
    \textit{MAE} $\downarrow$        & $\num{6.77e-13}$             & \multicolumn{1}{c|}{$\num{2.18e-13}$}  & $\mathbf{\num{1.93e-13}}$ \\ \hline
    \textit{RMSE} $\downarrow$       & $\num{1.52e-12}$             & \multicolumn{1}{c|}{$\num{4.73e-13}$}  & $\mathbf{\num{4.03e-13}}$     \\ \hline
    \textit{MAPE} $\downarrow$       & $75.7\%$                     & \multicolumn{1}{c|}{$\mathbf{24.2\%}$} & $28.9\%$                  \\ \hline
    \textit{S-MAPE} $\downarrow$     & $47.6\%$                     & \multicolumn{1}{c|}{$\mathbf{22.0\%}$} & $35.8\%$                  \\ \hline
    $\bar{\xi} \downarrow$             & $\num{6.77e-13}$             & \multicolumn{1}{c|}{$\num{2.18e-13}$}  & $\mathbf{\num{1.93e-13}}$ \\ \hline
    $\sigma_\xi \downarrow$            & $\num{1.35e-12}$             & \multicolumn{1}{c|}{$\num{3.67e-13}$}  & $\mathbf{\num{3.01e-13}}$ \\ \hline
    $\max_\xi \downarrow$              & $\num{1.30e-11}$             & \multicolumn{1}{c|}{$\num{5.17e-12}$}  & $\mathbf{\num{4.03e-12}}$ \\ \hline
    \end{tabular}
    \label{tab:results}
\end{table}

\begin{figure}[!ht]
    \centering
    \includeinkscape[width=0.9\linewidth]{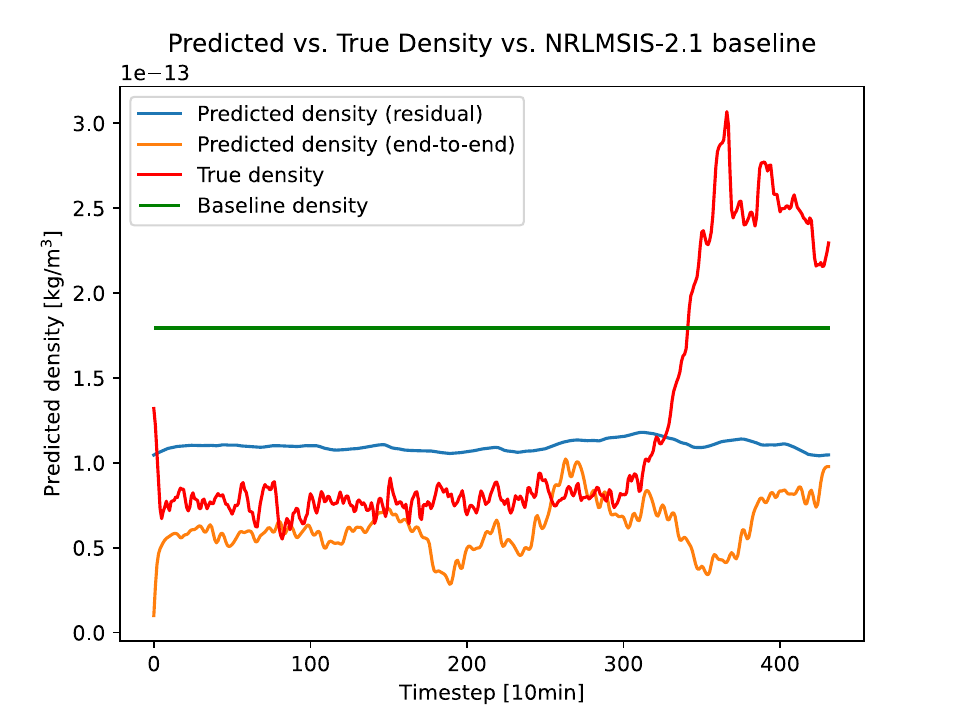}
    \vspace{-0.3cm} %
    \caption{
    Prediction of the thermospheric density using our proposed transformer with end-to-end and residual learning approaches compared to the persistence baseline and the measured density by SWARM-A (August 2019).
    All models, including the persistent baseline, are unable to capture the rise of the measured density.
    }
    \label{fig:boring-hallucinated}
\end{figure}

Despite the strong performance demonstrated in the aggregated metrics, there are cases in which Transformer models using both training schemes fail to accurately predict thermospheric density, as illustrated in \autoref{fig:boring-hallucinated}.
In this example, a spontaneous increase in solar activity, approximately two days into the prediction horizon, causes a sharp increase in the density.
None of the models are capable of anticipating this behavior, likely because the relevant precursors are absent from the input data, which presents an inherent flaw in simulating long prediction horizons.\vspace{-0.15cm}

\section{Discussion}
\label{sec:discussion}
\vspace{-0.3cm}
Despite the limited training set of $6153$ samples after the train/validation split, our proposed model demonstrates substantial improvement over the baseline, as detailed in \autoref{sec:results}.
As shown by \autoref{fig:captured-well}, especially but not limited to time horizons following recent solar events, the model predictions are notably more reliable.
This is further supported by the improved error metrics and remaining errors statistics.
In these instances, the persistence baseline fails to react to strong solar events shortly before the prediction horizon, wheras our model adapts more effectively. 
Training with the $\text{OD-RMSE-Loss}$ also proves to be effective, as especially early predictions are generally more reliable than later ones.
This is important as the error in regressive models gets propagated over the course of the next predictions.

However, the model's limitations become apparent during episodes of abrupt and extreme changes, particularly when spontaneous solar events occur within the prediction window.
Since these events are probably also not modeled by any of the input features, these deviations are inherently difficult to predict without additional specialized data.
Integrating such data would likely increase the model complexity and thus requires more training samples, especially data containing such events.
Another limitation lies in the model's tendency to overfit across many configurations, underscoring the need for more data.
With $6160$ samples used to train almost $1.8$ million tunable parameters ($1,795,585$), there is clear room for improvement in scaling.
We attempted to reduce model size but encountered degradation in predictive performance.
Up to this point, the impact of each input has also not yet been validated, which requires an ablation study optionally paired with an analysis of gradients.
An investigation of better-suited, available inputs that could increase the model's performance, due to more efficient knowledge transfer, is necessary.

Due to the strong improvements compared to the persistence model often used today, a real-world operation of this model for planning s/c maneuvers at this point is probably feasible under the assumption of prior precautions.
If there is an on-board sensor that measures the current density or the actual one is inferred from measured orbit perturbations, there is the possibility of capturing the uncertainty of the proposed model.
This allows for discarding the prediction if outliers are to be expected.

\section{Conclusion}
\label{sec:conclusion}
\vspace{-0.3cm}The proposed transformer is capable of beating the predictions of a persistence baseline model, is easy to run, and takes similar simple inputs.
The effectiveness of the model is shown, although we remark that in spontaneous occasions the predictions are off.
We are eager to improve our approach in the future by first increasing the amount of data being used to train the model or by testing approaches where we use physical models, like TIE-GCM, to generate artificial data.
With the scaling laws of Transformers in mind, big improvements are suspected with an increased dataset size.
Furthermore, we believe that uncertainty predictions obtained by training model ensembles are crucial for a deployment in mission planning, as this can help in planning for alternative scenarios.\vspace{-0.1cm}

\bibliography{ifacconf}

\end{document}